\documentstyle[aps,prl,multicol,epsf]{revtex}

\newcommand{\be}{\begin{equation}}
\newcommand{\ee}{\end{equation}}
\newcommand{\bea}{\begin{eqnarray}}
\newcommand{\eea}{\end{eqnarray}}
\newcommand{\bfig}{\begin{figure}}
\newcommand{\efig}{\end{figure}}
\newcommand{\bc}{\begin{center}}
\newcommand{\ec}{\end{center}}

\newcommand{\minf}{M_\infty}
\newcommand{\p}{\phi}
\newcommand{\pw}{\phi_w}
\newcommand{\szz}{\sigma_{zz}}
\newcommand{\sxx}{\sigma_{rr}}

\newcommand{\sxz}{\sigma_{rz}}
\newcommand{\snn}{\sigma_{nn}}

\newcommand{\snm}{\sigma_{nm}}

\newcommand{\sij}{\sigma_{ij}}

\begin{document}

\title{Stresses in silos: Comparison between theoretical models and
new experiments}

\author{L. Vanel$^1$, Ph. Claudin$^2$, J.-Ph. Bouchaud$^2$,
M.E. Cates$^3$, E. Cl\'ement$^1$, and J.P. Wittmer$^4$}

\address{$^1$ L.M.D.H., Univ. Paris VI, 4 place Jussieu - case 86,
75005 Paris, France.} \address{$^2$ Service de Physique de l'Etat
Condens\'e, CEA, Orme des Merisiers, 91191 Gif-sur-Yvette Cedex,
France.} \address{ $^3$ Dept. of Physics and Astronomy, Univ. of
Edinburgh, JCMB King's Buildings, Mayfield Road, Edinburgh EH9 3JZ,
GB.} \address{ $^4$ Dept. de Physique des Mat\'eriaux, Univ. C.
Bernard - Lyon I, 43 Bvd du 11 Novembre 1918, 69622 Villeurbanne,
France.}

\maketitle

\begin{abstract} We present precise and reproducible mean pressure
measurements at the bottom of a cylindrical granular column. If a
constant overload is added, the pressure is linear in overload and
nonmonotonic in the column height. The results are {\em
quantitatively} consistent with a local, linear relation between
stress components, as was recently proposed by some of us. They
contradict the simplest classical (Janssen) approximation, and may
pose a rather severe test of competing models.

\end{abstract}

\centerline{PACS numbers: 46.10.+z, 83.70.Fn}
\bigskip

\begin{multicols}{2}
\narrowtext

The prediction of static stresses in dry, cohesionless granular matter
has become the focus of renewed attention (see
\cite{deGennes,cargese,savage,proc}). Surprisingly, there is no
consensus on what is the basic physics involved. Some argue that the
behavior is essentially elastic (ultimately justified by the slight
elastic deformation of individual grains)\cite{deGennes}; others that
it is dominated by the extremely nonlinear constraint that tensile
intergranular forces are absent \cite{prl,moukarzel}. Indeed, some of
us \cite{proc,bcc,nature} have argued that the statics of granular
materials can be described, without considering elastic displacements,
by assuming a local, history-dependent, relation  between stress
tensor components. This gives {\em hyperbolic} equations for the
stress field, in contrast to the {\em elliptic} (or
elliptic-hyperbolic) equations of conventional elastic (or
elastoplastic) models.  Our approach provides a simple continuum model
of `force chains' \cite{dantu,radjai,EandO}; (physical) force chains
become (mathematical) characteristics of the hyperbolic equations. In
the simplest case, these form a regular array; stresses propagate
through space via a wave equation \cite{bcc,prl}. According to the
model, the medium is `fragile' in a precise sense \cite{prl}: it
responds linearly to a specific class of `compatible' loads; all
others cause plastic reorganization.

This approach accounts well \cite{nature} for the pressure `dip'
below the
apex of a conical sandpile poured from a point source \cite{Smid}. (It
also predicts \cite{bcc,nature} that the dip is {\em absent} for a
pile made of successive horizontal layers, as recently confirmed by
experiment \cite{Vanel}.) However, it has excited strong criticism in
some quarters \cite{savage}, and certainly demands further
experimental test \cite{deGennes}. For example, such models predict
that if a small localized overload is placed on top of a granular
layer, the excess weight at the bottom is  maximal, not directly
beneath the weight, but on a {\em ring} \cite{cargese,bcc}. To test
this directly is difficult, because of strong nonlinearity and
(especially) noise effects which hinder the interpretation of data
\cite{pre,EC}.

A more robust and practical situation, is the cylindrical granular
column, or bin. Here also noise effects come into play; but ways
around these (by careful ensemble averaging of experimental data) have
been pioneered in \cite{Vanelsilos}. Below we report precise
measurements (beyond those of \cite{Vanelsilos}) of the effective mass
$M_e$, supported by the bottom plate, as a function of the total mass
$M_t$ poured into a (small) bin, with and without an added overload.
With no overload, as expected, $M_e(M_t)$ first rises linearly, then
saturates at a column height comparable to its width; for high bins,
most of the mass is `screened' by frictional transfer to the walls. A
simple hyperbolic model (called {\sc osl} for `oriented stress
linearity' \cite{nature}) gives bin results close to, but different
from, the classical Janssen approximation (recalled below)
\cite{bcc,these}. In contrast to traditional methodologies
\cite{BandR} our new ensemble-averaged experiments can distinguish
these predictions; we find that {\sc osl}, which has an extra fitting
parameter, is discernibly better. Another classical model ({\sc ife},
see below) gives wholly inadequate answers unless unphysical values of
the wall and bulk friction constants are used.

There then follow, from the {\sc osl}
model, two
important new predictions for the effect of a uniform overload of mass
$Q$ at the top of the granular column. First, $M_e$ should be {\em
linear} in $Q$; second, for large $Q$, $M_e$ should be {\em
nonmonotonic} in $M_t$. We find that, with no further fitting, our
overload experiments {\em quantitatively} confirm the {\sc osl}
predictions, strongly supporting the hyperbolic picture. At the end of
this Letter, we comment on the challenge these new results pose to
other modelling strategies.

First we recall our own approach. By stress continuity,
\be
\nabla_{i}\sij = \rho g_j\label{silo2dequi1}
\ee  where $\sij$ is the (symmetric) stress tensor, $\rho$ is the
density of the material, and $g_j$ is the gravitational acceleration.
In general one needs extra physical assumptions to close
Eq.\ref{silo2dequi1}. For an elastic body, one assumes a
(single-valued) displacement field, and a linear relation between
stresses and strains (Hooke's law). For poured cohesionless grains,
the definition of a macroscopic displacement is problematic (see
\cite{deGennes,proc}). Instead we assume that the arrangement of
granular contacts gives, on continuum length scales, a definite
relation between components of the stress tensor
\cite{bcc,Grinev,moukarzel}. One such relation, often used in the
literature, is the {\sc ife} (`incipient failure everywhere')
assumption: that the material is everywhere on the verge of Coulombic
failure (see e.g. \cite{Nedderman,nature}). Then there exists a
(locally varying) set of axes ${\bf n}\perp{\bf m}$ such that $\snm =
\snn\,\tan \p$ where $\p$ is the Coulomb angle. 

Our modelling strategy instead gives a fundamental r\^ole to the
network of force chains which, if grains are undeformable, must carry
forces longitudinally \cite{prl}. One interpretation of our equations
is that the friction between parallel force chains is fully mobilized;
a Coulomb-like condition, $\snm = \snn\,\tan \psi$, then holds (with
$\psi \leq \p$ an `effective' friction angle) but the orientation
${\bf m}$, which is directed along the force chains, is now fixed by
the construction history and {\em not} (as in {\sc ife}) by the load
\cite{prl,these}. (This assumes the load is a compatible one.) For
simple construction histories, like piles and bins, we assume that
${\bf m}$ is the same everywhere, up to an inversion through the
central symmetry axis; ${\bf m}$ must then have a fixed angle $\tau$
to the vertical. In cylindrical polars ($z,r,\theta$) with $z$
downwards, we recover the {\sc osl} model \cite{nature}: \be
\label{silooslrel} \sxx = \eta_1 \szz + \eta_2 \sxz \ee with $\eta_1=\tan\tau
\ \cot(\tau-\psi)$ and $\eta_2=\tan\tau -\cot(\tau-\psi)$.
Eq.\ref{silooslrel} closes the problem in two dimensions ($d=2$):
inserting it into Eq.\ref{silo2dequi1}, gives an anisotropic wave
equation, with one characteristic along ${\bf m}$, and another along a
direction ${\bf m'}$  at angle $\tau-\psi-\pi/2$. (These can be
interchanged without affecting Eq.\ref{silooslrel};  so ${\bf m'}$ describes a second family of force chains \cite{prl}.) For $d=3$,
a further closure equation is needed. Our choice here is $\sigma_{rr} =
\sigma_{\theta\theta}$; but from work on conical piles, we expect
insensitivity to this choice \cite{nature,Nedderman}.  In the bin
geometry, the {\sc osl} model can then be solved exactly ($d=2$)
\cite{bcc,these} or numerically ($d=3$). Note that {\sc ife}, like
{\sc osl}, gives propagative (hyperbolic) equations; but these are
nonlinear, unlike our wave equation.

For nonzero $\eta_2$, the force chain network
distinguishes between inward and outward radial directions. 
This does not contradict the axial symmetry present \cite{nature}. But
if as well the medium is {\em locally} symmetric, then $\eta_2 = 0$; in
Eq.\ref{silooslrel}, this recovers the model of Ref.\cite{bcc}. The
latter can  be viewed as a local version of the classical Janssen
hypothesis \cite{Nedderman,Janssen}. Janssen proposed  a constant
ratio between  horizontal and vertical stresses, $\sxx = K \szz$, but
neglected altogether their dependence on $r$. Assuming also that
friction at the wall is fully mobilized, with a friction coefficient
$\tan \pw$, he found the equation: \be M_e = \minf \left(1 -
\exp\left[{-{M_t}/{\minf}}\right]\right) \label{janssen} \ee with
$\minf={\rho D^2}/{2K\tan\pw}$ for $d=2$, and $\minf= {\rho\pi
D^3}/{16K\tan\pw}$ for $d=3$; $D$ is the bin diameter.

We turn now to the experimental procedure, described in detail in
\cite{Vanelsilos}. The bin is a tube of diameter $D=3.8$ cm, filled
with beads of glass (density $\rho_b=2.6$ g/cm$^3$, diameter $2$ mm).
The bottom comprises a very stiff scale plate ($2\times 10^4$ N/m).
Initially, the tube is filled with a low packing density; this is
increased by giving it small taps. The bottom plate is then lowered
(by a few tens of microns) and the effective mass decreases
monotonically to an asymptotic value; $M_e$ and the mean density
$\rho$ are measured. The density is again increased by tapping, the
plate lowered and further measurements taken. This entire procedure is
done about $30$ times -- each run giving results for the whole range
of densities. The measured results for $M_e$ show  a certain
($M_t$-dependent) `error bar': not a measurement error of the mass,
but arising from intrinsic fluctuations in the packing. This protocol
is a major advance because (a) an ensemble average value for $M_e$ is
found, improving accuracy; (b) due to the downward motion of the base,
that wall friction is fully mobilized, which might not
otherwise be the case \cite{EdeG}. The wall friction angle is measured
separately as $\pw=22^o \pm 2^o$ \cite{Vanelsilos}, thus eliminating
one fit parameter.

\bfig[h]
\epsfxsize=8.6cm
\epsfbox{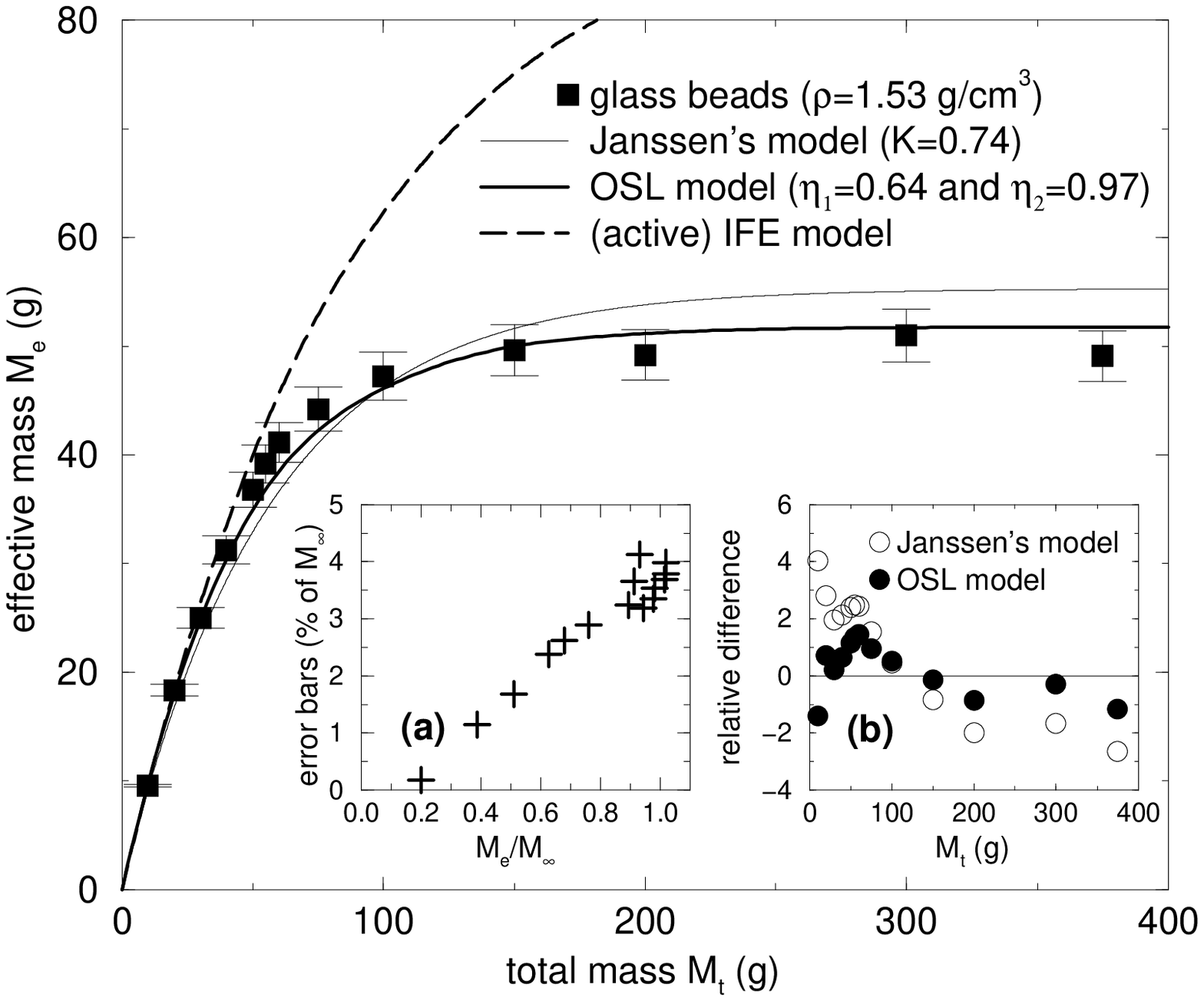}
\caption{Main figure: Experimental and theoretical $M_e(M_t)$ curves.
(a): Statistical dispersion of the measures. (b): Relative deviation
between experiment and theory, i.e. $\delta M_e/\Delta$.} \efig

The experimental results, for a packing density $\rho=1.53$ g/cm$^3$,
are compared in Fig.1 with three models: {\sc ife} (which has no
adjustable parameter once the internal friction angle $\p=25^o \pm
2^o$ is known); Janssen's equation (one adjustable parameter); and the
{\sc osl} model (two adjustable parameters). Each plotted datapoint is
itself a mean value, with an error bar $\Delta$ shown in inset (a).
(This is small at small $M_e$ but then grows rapidly.) To find the
best fits, we have minimized the following: \be \label{defE} E^2 =
N^{-1}\sum_i (\delta M^i_e/\Delta^i)^2 \ee where $\delta M^i_e$ is the
difference between the $i$th experimental datapoint and the
theoretical $M_e$ value, $\Delta^i$ the observed error bar, and $N$
the number of datapoints.

For our data, the (active) {\sc ife} approach, using the {\em measured} 
friction values $\phi$ and $\phi_w$ is plainly inadequate.
Better agreement with {\sc ife} is found by taking
$\phi$ and/or $\phi_w$ as fit parameters. Even then, the fit 
remains poor (e.g.
$E=4.43$ for $\rho=1.53$ g/cm$^3$) ; and the fitted
values, $\phi=\phi_w=30^o$ are {\em incompatible} with those found by
direct experiment. For given $\phi_w$, {\sc ife} systematically {\em
overpredicts} the asymptotic stress; so the fitted $\phi_w$ exceeds
the real one. In systems where the wall friction is not fully
mobilized, the error is harmlessly absorbed by the fit. In our system,
the fitted value is {\em higher} than the fully mobilized $\phi_w$
measured separately, which is unphysical.

Unlike the {\sc ife} model,
Janssen's model gives a fair approximation ($E \sim 2$; Table 1)  but,
as shown in inset (b),  there is a clear systematic deviation:
screening by the walls is in turn over- and underestimated for small
and large $M_t$ values. (Note also that our $K$ parameters are higher
than those usually reported \cite{BandR,Oin}: but as with {\sc ife}, low
fitted values might compensate for incompletely mobilized of wall
friction.) This has led two of us \cite{Vanelsilos} to propose
elsewhere an empirical model (not shown) where an excess contribution
from grains at the bottom of the pile is added to the Janssen result.
As shown in Table 1, the best-fit {\sc osl} model does as well as this
empirical model, with an error $E \sim 1$ \cite{RqQ}:  the systematic
deviations are reduced, in particular in the first part of the curve.
This can be understood by noting that within the {\sc osl} model, the
grains contained within a `light-cone', resting on the bottom plate,
cannot interact with the walls \cite{Vanelsilos}; the mass of these
grains  is completely unscreened.

\bfig
\bc
\begin{tabular}{|c|c|c|c|}
\hline
density              & {\sc ife} & Janssen       & {\sc osl}   \\
\hline
$\rho=1.51$ g/cm$^3$ & $E=5.96$ & $E=2.11$       & $E=0.89$    \\
($\rho/\rho_b=.58$)  &          & $\minf=61.9$ g & $\eta_1=0.55$ \\
       &                & K=0.65         & $\eta_2=1.03$  \\
\hline
$\rho=1.53$ g/cm$^3$ & $E=8.55$ & $E=2.28$       & $E=0.94$    \\
($\rho/\rho_b=.59$)  &          & $\minf=55.3$ g & $\eta_1=0.64$ \\
       &                & K=0.74         & $\eta_2=0.97$  \\
\hline
$\rho=1.56$ g/cm$^3$ & $E=10.1$ & $E=2.28$       & $E=1.02$    \\
($\rho/\rho_b=.60$)  &          & $\minf=52.3$ g & $\eta_1=0.71$ \\
       &                & K=0.80         & $\eta_2=0.85$  \\
\hline
$\rho=1.59$ g/cm$^3$ & $E=12.4$ & $E=2.30$       & $E=1.08$    \\
($\rho/\rho_b=.61$)  &          & $\minf=48.5$ g & $\eta_1=0.87$ \\
       &                & K=0.87         & $\eta_2=0.49$  \\
\hline
\end{tabular}
\ec

Table 1. {\small Results of the fits of the experimental data points,
and the corresponding physical parameters.} \efig

Note the values found for $\eta_2$. The minimum of
$E(\eta_2)$ is not sharp, but {\em positive} $\eta_2$ is always
preferred (as
for other types of grains
\cite{these}). A positive
$\eta_2$ means that most of the weight follows the `inward'
characteristic
thus reducing the screening effect of the walls. Conversely,
in sandpiles (created from a point source) $\eta_2$ is {\em negative}
\cite{nature,these}; this `outward' transfer of weight is responsible
for the pressure dip underneath the apex. Positive $\eta_2$ could be
caused by slight inward avalanches of material as the base is lowered.
Its decrease at higher densities might indicate a diminished
susceptibility to this effect; alternatively the tapping procedure
could progressively erase a local assymmetry induced by the initial
fill.

\bfig[h]
\epsfxsize=8.6cm
\epsfbox{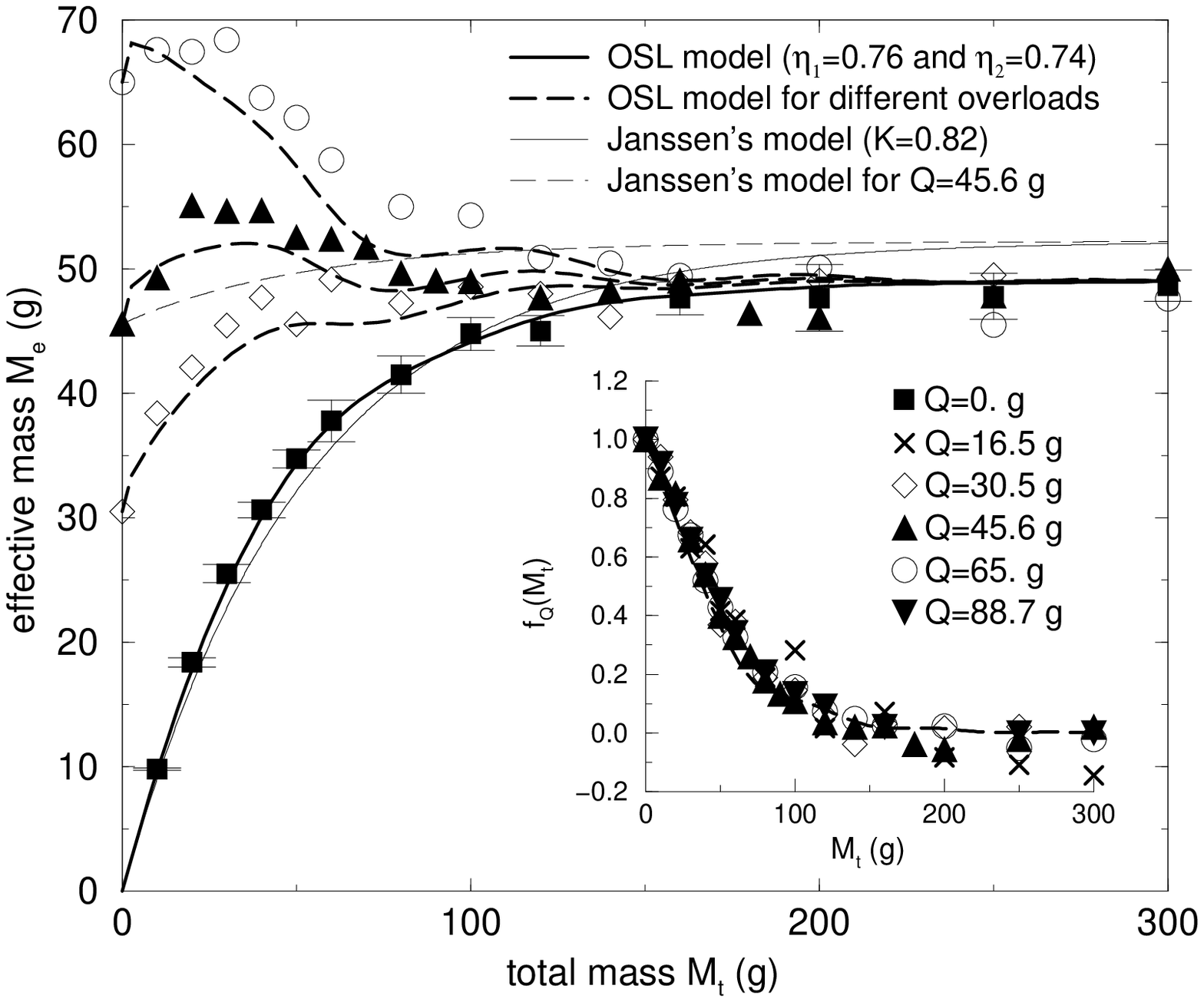}
\caption{$M_e(M_t)$ for various overloads
$Q$. Symbols: experiment
($\rho=1.60$ g/cm$^3$); long-dashed lines: {\sc osl}
predictions. Inset: $(M_e-\minf
f_0)/Q$ showing linear data collapse.}
\efig

We now turn to the key results of this paper, for the
response
to an overload
$Q$ placed on top of the granular column. (This is a solid
piston, just narrower than the cylinder.) This is taken into
account within
the {\sc osl} model by modifying the boundary conditions to
include a uniform downward stress at the top
surface.
Such a load is found to be compatible. From the linearity of the {\sc osl} model (also true of Janssen's
model) we then have: \be M_e = \minf f_0\left(\frac{M_t}{\minf}\right)
+ Q f_Q\left(\frac{M_t}{\minf}\right) \label{janssenbis} \ee In
Janssen's description, $f_0(x)=1-e^{-x}$ and $f_Q(x)=e^{-x}$, so $M_e$
is monotonic in the poured mass $M_t$ (and constant when $Q=\minf$).
The result of the {\sc osl} model are more surprising: $f_0(x)$ and
$f_Q(x)$ have different $x$ dependences. Hence $M_e$ is not monotonic
in $M_t$; at intermediate $Q$ there is an `overshoot' (Fig.2). In
addition, both functions have a (slight) oscillatory character, caused
by `resonances': these are standing-wave modes of the wave equation,
damped by `absorption' arising from wall friction (see \cite{bcc} and
\cite{these}). In Fig.2, we show the experimental results obtained for
various overloads $Q$. As shown in the inset, these results do indeed
obey the linear relation, Eq.\ref{janssenbis}, to good accuracy. A
clear overshoot effect is also seen, although any further `resonant'
oscillations are small (even theoretically). Note that the {\sc osl}
predictions in Fig.2 use the {\em same} parameters as determined
previously for $Q=0$. Thus {\sc osl}, with no further fitting, gives a
good quantitative account of the data for all $Q$.

We have shown that simple hyperbolic models \cite{bcc,nature},
encoding the presence of linear force chains \cite{proc,prl}, can be
used to reproduce {\em quantitatively} the observed stress response of
cohesionless granular media, not only in piles \cite{nature}, but in
bins. The same is not true of the traditional Janssen analysis. Nor is
it true of {\sc ife}; this does predict resonant behaviour (at least
in local stresses \cite{Nedderman}), but our results, even without
overload, rule it out entirely as a physical model. Any expectation of
nonmonotonicity in Fig.2 based on {\sc ife} would thus have been
misplaced.

What of other continuum modelling strategies?
Much  recent work on bins and silos has studied
elastoplastic constitutive models (also widespread in soil mechanics),
often by a finite-element method. There are many such models, and a
recent comparative study found little consensus among them
\cite{rotter}.  But we wonder whether these approaches can, with
reasonably few fit parameters, reproduce the results of Figs.1 and 2.
For example, the observed linearity in $Q$ (seen even for
$Q/\minf\simeq 1$) may set a challenge, although one finds numerically
that, after summing stresses over the base, the (non-linear) {\sc ife}
model obeys to a good precision the linear relation
(\ref{janssenbis}).
Linearity is, of course, also recovered if the material is entirely
Hookean. The challenge is then to explain within a purely elastic
theory the nonmonotonic (if not oscillatory) curves of Fig.2. The
investigation of these important questions is underway \cite{Roux}.

Finally it is important to map out more clearly the domain of validity
of the hyperbolic approach (see e.g. \cite{Grinev}). In particular,
our granular columns are tiny: only twenty grains or so across. These
data clearly do not rule out a crossover to more conventional elastic
or elastoplastic behavior at larger scales (e.g. where the grains
start to deform) \cite{proc,prl}, although the hyperbolic approach
also works well in conical piles up to 1 metre wide \cite{Smid}. 
Careful overload experiments on larger bins could be very valuable, as
well as local stress measurements, which would reveal more clearly the
oscillatory nature of the response.

We thank P.G. de Gennes, J. N. Roux and G. Combe for very useful discussions.
E.C. and L.V. thank J. Lanuza for technical assistance.

\end{multicols}
\end{document}